\documentclass[preprint]{aastex}

\shorttitle{Intracluster PNe in Virgo}
\shortauthors{Arnaboldi, M., Aguerri, J. A. L. et al.}

\begin{document}


\title{Intracluster Planetary Nebulae in Virgo: Photometric
selection, spectroscopic validation and cluster depth
\footnote{This paper is based on observations carried out at the 
ESO telescopes at La Silla and at the Anglo Australian Observatory.}}


\author{Magda Arnaboldi}
\affil{Astronomical Observatory of Capodimonte, Via Moiariello 16, 
I-80131 Naples}
\email{magda@na.astro.it}

\author{J. Alfonso L.  Aguerri}
\affil{Astronomisches Institut der Universit\"at Basel, Venusstrasse 7, 
Binningen, Switzerland}
\email{jalfonso@astro.unibas.ch}

\author{Nicola R. Napolitano}
\affil{Astronomical Observatory of Capodimonte, Via Moiariello 16, 
I-80131 Naples\\
Dept. of Physics, University "Federico II'', I-80125 Naples}
\email{napolita@na.astro.it}

\author{Ortwin Gerhard}
\affil{Astronomisches Institut der Universit\"at Basel, Venusstrasse 7, 
Binningen, Switzerland}
\email{gerhard@astro.unibas.ch}

\author{Kenneth C. Freeman}
\affil{RSAA, Mt. Stromlo Observatory, Weston Creek P.O., ACT 2611}
\email{kcf@mso.anu.edu.au}

\author{John Feldmeier}
\affil{Case Western Reserve University, 10900 Euclid Avenue, Cleveland, 
OH 44106-7215}
\email{johnf@eor.astr.cwru.edu}

\author{Massimo Capaccioli}
\affil{Astronomical Observatory of Capodimonte, 
Via Moiariello 16, I-80131 Naples\\
Dept. of Physics, University ``Federico II'', I-80125 Naples}
\email{capaccioli@na.astro.it}

\author{Rolf P. Kudritzki}
\affil{Institute for Astronomy, Honolulu, Hawaii 96822 USA}
\email{kud@ifa.hawaii.edu}

\author{Roberto H. M\'endez}
\affil{Munich University Observatory, Scheinerstr. 1, D-81679 Munich}
\email{mendez@usm.uni-muenchen.de}



\begin{abstract}
We have imaged an empty area of $34\times34$ arcmin$^2$ one and a half
degree north of the Virgo cluster core to survey for intracluster
planetary nebula candidates. We have implemented and tested a fully
automatic procedure for the selection of emission line objects in
wide-field images, based on the on-off technique from Ciardullo and
Jacoby.  Freeman et al. have spectroscopically confirmed a sample of
intracluster planetary nebulae in one Virgo field. We use the
photometric and morphological properties of this sample to test our
selection procedure. In our newly surveyed Virgo field, 75 objects
were identified as best candidates for intracluster PNe.

The luminosity function of the spectroscopically confirmed PNe shows a
brighter cut-off than the planetary nebula luminosity function for the
inner regions of M87. Such a brighter cut-off is also observed in the
newly surveyed field and indicates a smaller distance modulus,
implying that the front end of the Virgo cluster is closer to us by a
significant amount: 14\% closer (2.1 Mpc) than M87 for the 
spectroscopic field, using the PN luminosity function distance of 14.9 Mpc 
to M87,  and 19\% closer (2.8 Mpc) than M87 for the newly surveyed field. 
Independent distance indicators (Tully-Fisher
relation for Virgo spirals and surface brightness fluctuations for
Virgo ellipticals) agree with these findings.

From these two Virgo cluster fields there is no evidence that the
surface luminosity density for the diffuse stellar component in the
cluster decreases with radius.  The luminosity surface density of
the diffuse stellar population is comparable to that of the galaxies. 
\end{abstract}




\section{Introduction}
The presence of a diffuse intracluster light filling the regions
between galaxies in clusters has been known for several decades.  In
his introduction to the study of the galaxy luminosity function in the
Coma cluster in 1951, Zwicky suggested that ``... Vast and often very
irregular swarms of stars and other matter exist in the spaces between
the conventional spiral, elliptical, and irregular galaxies.  It is not
at all clear how such swarms should be incorporated into any
distribution function of galaxies..'' \cite{JL1.2}.  Early studies of
the diffuse light in clusters were based on photographic techniques
\citep{welch71,mel77,thuan77}, and were followed in
the 1990s by CCD observations \citep{guld89,uson91,ber95,vic94}. 
The limitations of photographic photometry
and the small CCD fields used in these works lead to large
uncertainties in the results.  The best estimates for the contribution
of the diffuse light ranged from 10\% to 50\% of the total light
emitted by cluster galaxies in the central regions of the clusters.
Such studies, which need accurate photometry on large sky areas,
encountered two problems: 1) the typical surface brightness of the
intracluster light is less than 1\% of the sky brightness and 2) it is
difficult to distinguish between the diffuse light associated with the
halo of the cD galaxy and that associated with the cluster
\cite{ber95}.

A different method for probing intracluster light is through the
direct detection of the stars themselves.  In 1977 a search was done
for intergalactic supernovae in the Coma cluster \cite{crtaw77}, but
all the supernovae detected were found to be associated with Coma
galaxies. In 1981 a supernova Type Ia was discovered in the region
between M86 and M84 in the Virgo cluster \cite{sm81}.  

A new approach came from the discovery of intracluster planetary nebulae 
(ICPNe) in the Virgo and Fornax clusters.  In a radial velocity survey of 
planetary nebulae (PNe) in NGC 4406 (v$_{sys} = -227$ kms$^{-1}$), 
three objects were found with radial velocities $v > 1300$ kms$^{-1}$ 
\cite{aa96}.
Theuns \& Warren (1997) found 10 ICPNe candidates in the Fornax cluster
and Ferguson et al. (1998) used HST to detect a statistical excess of
red stars in a field near M87 relative to the Hubble Deep Field; for follow-up
studies of intracluster light in Virgo see also Durrell et al. (2001).
Further surveys in Virgo \citep{me97,cia98,fel98,fel00} 
in the [OIII] $\lambda$ 5007
line resulted in the identification of more than a hundred ICPNe
candidates in the Virgo core region.

The first spectroscopic follow-up \citep{kud2000,kcf2000}
carried out on subsamples of these catalogues showed that the on-off
technique plus the ``by-eye'' identification of the PN candidates 
included some stars which were mis-classified as
emission line objects. In addition, because of the very faint limiting
fluxes reached by these surveys, they also detect high redshift
emission-line objects, most probably from Ly$\alpha$ emitters at
redshift $3.13$ and [OII] emitters at redshift 0.347.

We have started a survey project which aims at detecting ICPNe in
large areas in the Virgo cluster and tracing their number density
profile from the Virgo cluster center.  The large CCD mosaic images
used in this project require the implementation of a robust automatic
procedure to reliably select a population of emission line candidates.
In this paper we present the survey carried out in one field at an
angular distance of 1.5 deg from the Virgo core \cite{bin87}, and
discuss in some detail the selection criteria adopted for the
identification of ICPNe. A critical step in this procedure is the
validation against a spectroscopically confirmed sample of ICPNe
obtained by Freeeman et al.\ (2000, 2001 in prep.).

The survey of ICPNe in Virgo associated with the diffuse stellar
population is particularly relevant in the ongoing discussion for the
3D shape of the Virgo cluster.  The planetary nebulae luminosity
function (PNLF) has been extensively used as distance indicator for
early-type galaxies in Virgo and Fornax \citep{jac90,mc93,me2001}.
Therefore the ICPNe luminosity function can be a useful trace of the
Virgo cluster structure. Other distance indicators have advocated a
significant depth for the Virgo cluster.  The Tully-Fisher relation
\citep{pietu88,tal1990,yasu97} supports a distance range of 12 to 20 Mpc 
for the Virgo spirals. Surface brightness fluctuations for early-type 
galaxies \citep{neil2000,west2000} place most of the elliptical
galaxies at $\sim \pm 2$ Mpc around M87, but indicate a significant
fraction of background and foreground objects.  Surveys of ICPNe
covering large areas in the Virgo cluster would detect preferentially
those ICPNe at the closer edge of the Virgo cluster, and therefore
trace its extension quite efficiently.

In the following sections, we will review and discuss previous
procedures implemented for the selection of PNe in galaxies and empty
fields and test a method for the automatic selection of a reliable
ICPNe catalogue.  In Section~2 we present the new observation of a
Virgo cluster field covering $34'\times34'$ and the data reduction of
the mosaiced data. In Section~3 we present the extraction procedure
and test its completeness.  In Section~4 we discuss the candidate
selection from the extracted catalogue based on their morphology and
colors and test them against the spectroscopic validations.  In
Section~5 we discussed the properties, the luminosity function and the
inferred luminosity density in the different fields. Conclusions are
given in Section~6.

\section{Imaging observations and data processing}
\subsection{Observations}
In March 1999 we observed a field in the Virgo cluster $1.5$ 
degrees north  of the cluster core \cite{bin87} at coordinates 
$\alpha (J2000) = 12:26:12.8$ and
$\delta (J2000)= +14:08:02.7$; we will refer to this field as 
the RCN1 field. This field was selected in a region 
without bright galaxies, to target the intracluster population. 
We used the WFI on the ESO 2.2-m telescope to image this field through
an 80 \AA\ wide filter centred
at 5023 \AA, which is the wavelength of the [OIII] $\lambda$ 5007 
\AA\ emission at the redshift of Virgo cluster (the ``on-band'' filter), 
plus a broad V band (the ``off-band'' filter). The total exposure times
were 24000s
in the narrow band filter, and 2400s in the V band. Individual frame
exposure times were 3000s for the narrow band filter and
300s for the V broad band, and were acquired following an elongated
rhombi-like dithering pattern in order to remove the CCD gaps, bad pixels and
hot/cold columns, plus optimum flat fielding.
Both sets of images were taken under similar 
seeing conditions ($\approx 1.2 ''$). 

Our new field covers a much larger area than 
previous observations for the detection of ICPNe 
in Virgo fields \citep{fel98,me97,aa96}. 
The whole image consists of a mosaic of $4 \times 2$ CCD 
detectors with narrow inter-chip gaps. One pointing  of the WFI covers 
34$'\times$ 34$'$.
The WFI combines a large field of view with 
a small pixel size of $0.238''$. The CCDs have an average
read out noise of 4.5 ADU/pixel and a gain of 2.2 e$^-$/ADU.
During the night, several Landolt fields and spectrophotometric
standard stars were acquired for the broad band and narrow band flux 
calibrations.

\subsection{Data reduction}
Here we summarise the basic steps for the data reduction: for a detailed
treatment of the mosaic data 
we refer the reader to Capaccioli et al. (2001).
The removal of instrumental signature (bias subtraction and 
flat-field corrections) was performed using the tasks in the 
MSCRED package in IRAF.  
The electronic bias level was removed from each CCD by 
fitting a Chebyshev function to the overscan region and subtracting it
from each column. 
By averaging 10 bias frames, a master bias was 
created and subtracted from the images in order to remove any remaining 
bias structures. The dark signal of each CCD detector was also 
subtracted from the images with the longest exposure time. 
Two types of flat-fields were taken: twilight and dome flats. 
Twilight flats were used for removing the large scale structure 
and the dome flats for the residual pixel-to-pixel structure.

Some structures were still present in the sky background at the
edges of the field, even after the twilight flat correction: 
this was additionally removed using a dark-sky super-flat image.
In CCD mosaiced data, superflats are needed because the optics
of the wide field imager diffuse light in a different way in twilight flats 
than in dark-sky images. 
The superflat was constructed as the median of the un-registered reduced 
dark-sky science images for each filter independently, with a $3\times\sigma$ 
rejection algorithm.
The residual structures in the background of the scientific frames are
corrected when this superflat is used\footnote{Because of the large
dithering pattern adopted during image acquisition, any diffuse light 
structures like ripples, shells or tidal tails as in 
Calc\'aneo-Rod\'an et al. (2000) would be averaged out in the 
superflat.}.\\
The final images are produced by stacking the dithered set of 
exposures obtained for each filter. 
Each reduced exposure must be corrected for the geometric distortions,
registered onto a common spatial reference, sky-subtracted and 
transparency-corrected before the final image can be produced. 
For further details, see the MSCRED package in IRAF and Capaccioli 
et al. (2001).

\subsection{Flux calibrations}
The strongest emission of a planetary nebula (PN) is frequently the 
[OIII] $\lambda 5007$\AA\ line \cite{dopi92}. 
The integrated flux from the [OIII] line of a PN is usually
expressed in a ``m(5007)'' magnitude using the definition
first introduced by Jacoby (1989):
\begin{equation}
m(5007) = -2.5 \log F_{5007} - 13.74
\end{equation}
which gives the PN luminosity in equivalent V magnitude\footnote{This 
definition approximates the apparent V magnitude one would see if all the 
[OIII] emission were distributed over the V bandpass.}
and $F_{5007}$ is the total [OIII] $\lambda$5007 
flux in erg cm$^{-2}$ s$^{-1}$ of each PN.
We have adopted a different system for the 
narrow band and broad band fluxes. To compare the limiting fluxes 
in the on-band and off-band, we normalise fluxes to
AB magnitudes, following Theuns \& Warren (1997).\\
The AB magnitude system is normalised to the Vega
flux as follows \cite{vega}:
\begin{equation}
m_{AB} = -2.5 (\log F_{\nu} + 19.436).
\end{equation}
The zero points for the narrow and broad band filters in the AB magnitude 
system were set from observations of the spectroscopic standard stars 
LTT 7379 and Hilt600. These two stars were offset to each CCD of the mosaic, 
and the errors on the zero points were determined from the RMS 
of the zero points from the 8 CCD of the mosaic. Several Landolt fields 
were observed for the calibration of the V band image.
The zero points for the [OIII] and V band imaging in AB magnitude are
$Z_{[OIII]} = 21.08 \pm 0.08$ and $Z_{V} = 23.76 \pm 0.08$, normalised 
to one second exposure. The error of $0.08$ on the zero point
is dominated by a non-uniform illumination of the WFI mosaic
(see Capaccioli et al. 2001 for a detailed discussion of this
problem). 
Airmasses were 1.3 and 1.2 for the reference [OIII] and the V band 
exposure\footnote{The reference [OIII] and V band exposures are 
those exposures in the dithered set of images which had the lowest
airmass and best seeing.}
respectively, and we have adopted the average extinction coefficient
for the La Silla observatory of $X_{[OIII]} = 0.16$ at 
$\lambda = 5030$\AA\ and of $X_{V} = 0.13$ 
for the V band.\\

We have computed the relation between the $m(5007)$ and the $m_{AB}$
magnitudes for the narrow band adopted filter for which
\begin{equation}
m(5007) = m_{AB} + 2.51;
\end{equation}
additional details on this calibration are found in Appendix A.
By using the AB magnitude system, the ``on-off'' band techniques
can be translated to a color selection criterium for an automatically 
generated catalogue.
Aperture magnitudes are computed in a fixed aperture of $3''$ diameter
(see discussion Sect.~3.1) for all objects, while the Kron magnitudes 
(mag(auto) in SExtractor) are computed in an aperture with radius equal 
to $2.5 \times$ the Kron radius \cite{kron80}, which scales with the
FWHM of the object luminosity profile.
The catalogues extracted in our field provide both aperture and Kron 
magnitudes for the [OIII] and V band images; the correction between 
aperture and Kron magnitude is 0.1 in both bands for stellar sources.  

\section{Extraction of emission line candidates}
Because of the strong [OIII] $\lambda 5007$\AA\ emission from PNe, 
they have so far been easily identified via the so called ``on-off" band 
technique: if the field of interest is imaged through a narrow band filter
centred at the redshifted [OIII] emission of a PN, and a second image 
is acquired using an off-band filter,
PNe candidates are usually identified as those point-like objects 
which are seen in the on-band image, and not seen in the off-band, 
when the two images are blinked. 
This ``on-off'' band technique and the ``by-eye'' identification 
of suitable candidates were used for PNe identifications 
in Virgo and Fornax ellipticals \citep{jac90,mc93,cia98}
and in empty intracluster fields \citep{me97,fel98}. 
However, spectroscopic follow-up of ICPNe candidates from 
\cite{fel98} has shown some contaminations from stars 
\cite{kud2000}, which may be caused by an off band image which is not ``deep'' 
enough.
A more quantitative color selection with an analysis of the limiting 
magnitudes in each of the two bands will allow a more detailed analysis of the 
completeness and contamination of the survey.

A survey of large fields requires an automatic procedure for the
candidate identification:  wide fields and high angular resolution,
with image frames of about 64 million pixels, makes the ``by-eye''
identification on blinking mosaic images difficult and time consuming.
The goal of this Section is to implement an automatic detection
algorithm, identify a series of optimal selection criteria and apply
such a procedure to our newly surveyed RCN1 field.  In the next
Section, the procedure will be tested against a spectroscopic confirmed
ICPNe sample from Freeman et al. (2000, 2001 in prep).

The approach by Theuns \& Warren (1997) is readily automated for large 
format CCD images. They selected IPNe candidates by their color excess in
$m_{n}-m_{v}$ color, where $m_{n}$ is the magnitude of the object in
the narrow band filter centred on the line emission, and $m_{v}$ is the
V band, both on in the AB magnitude system\footnote{Our off-band
filter includes the [OIII] line in its band pass, but we can neglect
this contribution because it would be 1.5 magnitudes below our V band 
detection limit even for the brighter objects}. Their best candidates are
those with a significant color excess in the narrow-band filter, i.e.
$m_{n}-m_{v}<<0$. This approach can be tested against spectroscopically
validated samples.

\subsection{Automatic candidate detections\label{sec3}}
We used SExtractor \cite{ber96} for the photometry in the broad V and
narrow bands. SExtractor identifies and measures fluxes from point-like
and extended objects in large format astronomical images.  Here we
discuss those parameters in SExtractor that are relevant for the
identification and flux measurements of our objects; for a detailed
description of the software we refer to Bertin \& Arnouts (1996).

Given a set of parameters for the source detection, 
we perform a set of tests to determine 
1) the completeness and 2) the fraction of spurious objects.
The parameter set for the source extraction will be the one 
for which the best compromise will be achieved.
We use a catalogue of simulated objects, homogeneously distributed in
our field with a given luminosity function, add these simulated objects 
to the narrow band image, and then look for the best parameter set
for the source detection which allows to reach the faintest flux for 
the modelled objects.
The fraction of spurious detections is derived from a simulated 
image which includes a realistic background plus the modelled 
catalogue\footnote{In this work the optimum
parameter set for extracting catalogues was obtained through a 
complete image simulation, i.e. background plus modeled catalogue.
The background was produced using the interpolated background image
and noise from the background-rms map, both produced by SExtractor 
as check-images (see SExtractor manual for additional details). 
To this background simulated image, we added the modeled catalogue of 
point-like sources.}.
An optimal parameter set for a source detection with SExtractor is the one 
which detects the largest number of real objects and produces the smallest
fraction of spurious detections, in the last magnitude bin.
The fraction of spurious detections depends strongly on the threshold
set for source detection and the image noise structure.

\begin{figure}
\epsscale{0.8}
\plotone{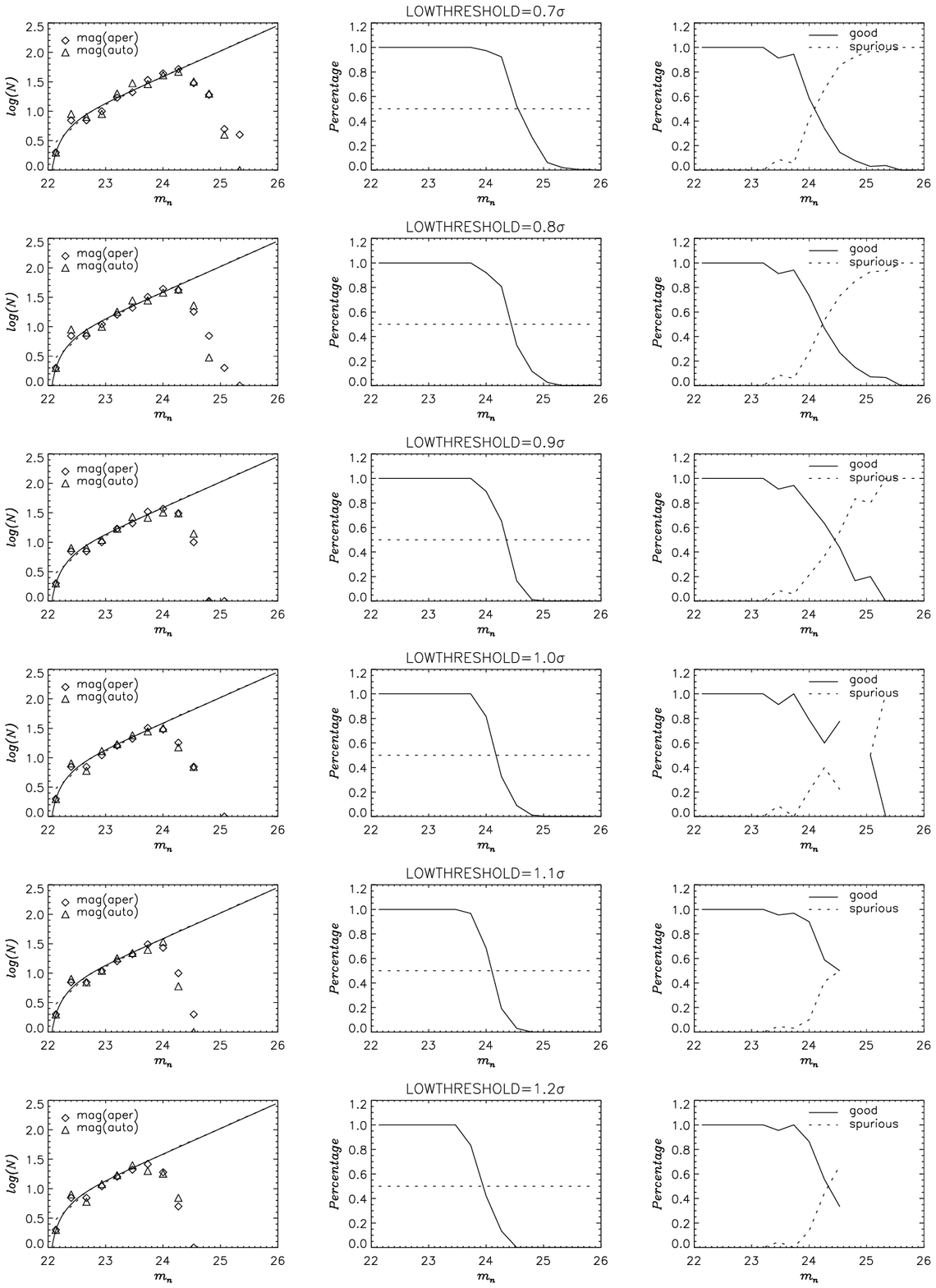}
\caption{Monte Carlo simulations carried out for a modelled population
of point-like objects.
Left panels: the luminosity function adopted for the modelled objects 
(continuous line) and the recovered luminosity function 
(triangles-mag(auto) and diamonds-mag(aper)). 
Central panels: Percentage of modelled point-like objects recovered for the 
different detection thresholds used. The dotted line at 50\% marks
the completeness limit.
Right panels: Percentage of the modelled point-like objects (continuous line) 
and of spurious objects
(dashed line) normalised to the total extracted object catalogue in 0.25
magnitude bins.
\label{fig1}}
\end{figure}

\begin{figure}
\plotone{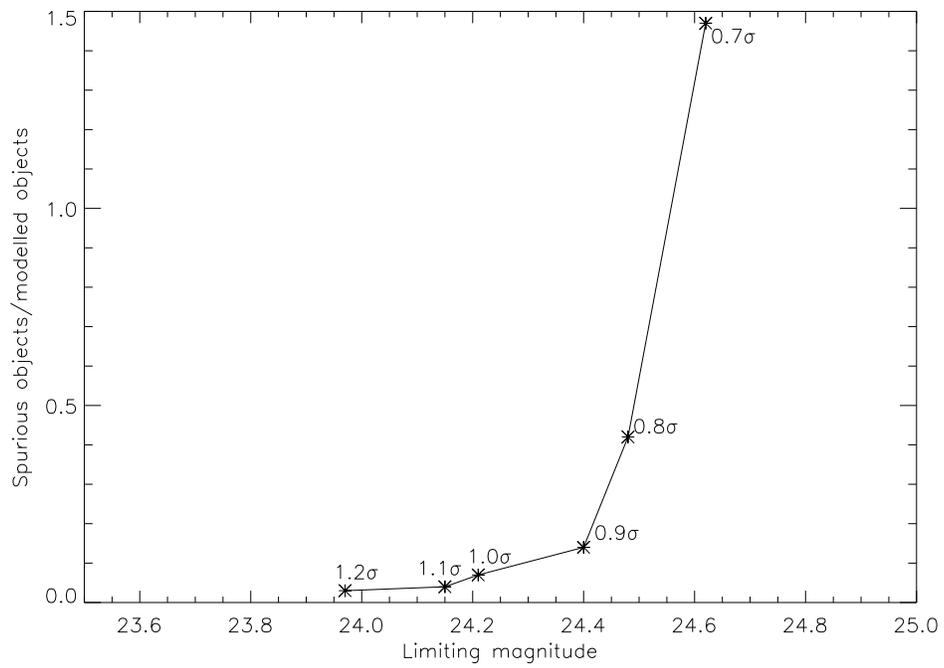}
\caption{Ratios of spurious to real detections plotted as a function of 
limiting magnitude for different low thresholds.\label{fig1b}}
\end{figure}

The simulated catalogue is computed from a luminosity function (LF)
which well reproduces the PNLF for the sample in M31,
see Eq.~(\ref{PNLF}), in the [OIII] AB magnitude interval $(22 \div 26)$. 
We tried detection thresholds between 0.7$\sigma$ and 1.2$\sigma$, 
where $\sigma$ is the global background RMS.
Table~1 gives the number of the retrieved 
sources from the modelled catalogue in magnitude bins and 
the number of spurious detections.
This number increases rapidly when the detection threshold is lowered,
see Figure~\ref{fig1} for a summary of the different tests.
On the other hand, by selecting a higher threshold, the number of 
extracted real sources decreases.
A compromise value for the detection threshold on our narrow band image 
is 0.9$\sigma$, because the differential decrease of the number of retrieved
real objects for detection threshold above 0.9$\sigma$ 
is always larger than the increase in the number of spurious detections,
as it is evident from Table~1 and Figure~\ref{fig1b}.
For the source extraction in the on and off-band
images, we decided then to use a detection threshold of $1.0\sigma$, 
to be conservative.

We can compute the limiting magnitude for the catalogue
extracted from our narrow band image as the magnitude 
for which 50$\%$ of the input objects are retrieved; 
the limiting magnitude is 24.2 with a $1.0\sigma$ threshold.
The estimated line fluxes for ICPNe in
Virgo cluster empty fields are in the range $25.8 < m(5007) < 27.5$
(Feldmeier et al. 1998). This magnitude interval corresponds
to $23.3 < m_{AB} < 25.0$ and thus our limiting magnitude
allows us to detect the first magnitude interval of the ICPNe 
luminosity function.

Emission line objects will be extracted from a master catalogue produced 
as follows:
\begin{itemize}
\item detect and measure magnitudes in a fixed aperture\footnote{For aperture 
photometry, we set a radius equal to $1.5'' = 3\sigma$, where 
$\sigma=\frac{FWHM}{2.356}$.  If a detected source is point-like, 
a $6\sigma$ diameter aperture contains 98.89\% of the total flux.}
for all objects in the [OIII] narrow band image, 
\item perform aperture photometry in the V band image 
at the (x,y) positions of the detected [OIII] sources, with
SExtractor in dual image mode (see SExtractor manual for additional details). 
\end{itemize}
The narrow band and V band image are registered before catalogue extraction.
The ICPNe candidates are identified as those point-like objects 
with $m_{n}-m_{v} << 0$ and no detected continuum. 
The objects with no detected continuum are those whose measured 
V mag(aper) values lie below the $1\sigma$ threshold
above sky; for our RCN1 V band image this corresponds to
24.75 mag. 

\section{Catalogues of emission line objects and spectroscopic validation}
\begin{figure}
\epsscale{0.7}
\plotone{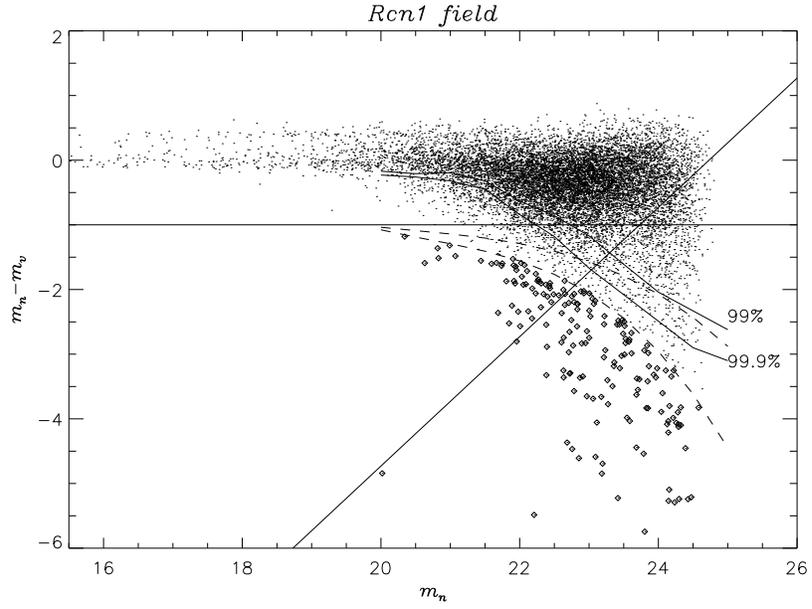}
\caption{Color-magnitude diagram for all the sources in the RCN1 field. 
The horizontal line at $m_{n}-m_{v}=-1$ indicates objects with an
observed EW=110 \AA. The diagonal line shows the magnitude corresponding to
$1.0 \times \sigma$ above sky in the V band. Full curved 
lines represent the 99$\%$ and 99.9$\%$ lines for the distribution of modelled 
continuum objects. The dashed lines represent 84\% and 97.5\% lines for the
distribution of modelled objects with $m_n - m_v=-1$. The points are all 
objects detected by SExtractor. Diamonds are objects with
a significant color excess in the narrow band filter. See text for details.
\label{fig2}}
\end{figure}

We extracted a catalogue of objects in the [OIII] image field RCN1 as 
described in Section~\ref{sec3}, for which we have aperture and Kron 
magnitudes in [OIII] ($m_n$) and V band ($m_v$). 
Figure~\ref{fig2} shows the color magnitude diagram (CMD) 
for the 18178 objects detected in our field. 
Most of them are continuum objects with an average $m_n - m_v$ color 
around $0$. There is some intrinsic scatter in the CMD, even with the two
filters being so close in wavelength: the sequence of blue and red stars can be
seen in the magnitude interval $m_n = 16 \div 22$.
The scatter in the colors of continuum objects at fainter magnitudes is
caused by two factors: contaminations by line-emitters and background field 
galaxies plus errors in the photometry for faint objects. 
Moreover most faint galaxies are red, and scatter up in the diagram;
we are interested in emission line objects whose color in $m_n - m_v$ 
is negative. ICPNe candidates are not the only emission line
objects with this property.
This color criteriom will in principle select at least the following 
four different kinds of emission line objects:
ICPNe and HII regions in Virgo,
[OII] $\lambda = 3726.9$ \AA\ emitters at z$\approx 0.34$,
Ly$\alpha$ galaxies at z$\approx 3.1$.
The contamination from [OII] emitters can be greatly reduced by considering 
only objects with observed equivalent width (EW) greater than 
100 \AA. 
Colless et al. (1990), and recently Hammer et al. (1997) and 
Hogg et al. (1998) found no [OII] emitters at z=0.35 with rest frame 
EW greater than 70 \AA, which implies an observed EW of 95 \AA.
Following Teplitz et al. (2000), if the emission line is a negligible 
contribution to the broad band flux, as it is in this case, 
the observed equivalenth width is
\begin{equation}
\mbox{EW}_{obs} \sim \Delta \lambda_{nb} ( 10^{0.4\Delta m} -1)
\end{equation}
where $\Delta \lambda_{nb}$ is the width of the narrow band filter in \AA.
Objects with observed EW$=95$\AA\ are those with color 
$m_{n}-m_{v} = -0.91$: our emission line candidates will be selected 
according to the condition $m_{n}-m_{v} = -1.0$ corresponding to an 
observed EW$= 110$\AA.

Due to the photometric errors at faint magnitudes, objects with intrinsic 
$m_{n}- m_{v} < -1.0$ do not all fall above the corresponding
straight line in the CMD. The scatter at faint magnitudes can mimic 
emission line objects from faint continuum sources.

To determine the effects of these photometric errors quantitatively,
we modelled a population of point-like sources with intrinsic colors
$m_{n}-m_{v}=0$ and -1,
and studied their distribution in the CMD. 
The continuous lines over plotted in Figure~\ref{fig2} represent the 
99$\%$ and 99.9$\%$ lines for a distribution of modelled objects with color 
$m_{n}-m_{v}=0$. Thus, above these lines will fall 99$\%$ and 99.9$\%$ 
of all continuum objects.
The dashed lines in Figure~\ref{fig2} 
represent the $1\times$RMS and $2\times$RMS lines of the 
color distribution for objects with $m_{n}-m_{v}=-1$. 
If the scatter of these objects in the CMD plane is Gaussian, 
then 84\% and 97.5\% of these objects are located above these lines. 

The ``on-off'' techniques for the identifications of
ICPNe requires that there be no detected emission in the off-band image,
which therefore needs to reach to fainter limiting AB magnitude to avoid 
contamination of our sample by faint continuum objects. 
The diagonal line in Figure~\ref{fig2} represents the V magnitude
for fluxes corresponding to $1.0 \times \sigma$ above sky in the
off-band. Objects above this diagonal line display some emission in V.
Objects whose m$_v$ fall below this line are not
detected by SExtractor\footnote{Their magnitudes are computed from
aperture photometry in a $3''$ diameter circular aperture at the positions 
of the [OIII] sources.} on the
off-band RCN1 image, with the selected low-threshold for detection.
Thus ICPNe candidates must be objects located in this region.
SExtractor in dual image mode cannot compute an $m_v$ magnitude (either 
mag(aper) or mag(auto)) for some of the 
[OIII] detected sources: these sources will all be considered as emission
line candidates.

\subsection{Point-like and extended objects}
The typical size of a PN is about $0.1 - 1$ pc; therefore
at the distance of the Virgo cluster ($\sim 15$ Mpc) they cannot be 
resolved on our images.  Two methods were used to
test whether an object is extended or point-like: 
(1) a 2D Gaussian fit and comparison to the PSF and (2) comparison between 
mag(auto) and mag(aper) magnitudes from SExtractor (see Section~\ref{sec3}). 

\subsubsection{Point-like objects: 2D Gaussian fit}
\begin{figure}
\epsscale{0.7}
\plotone{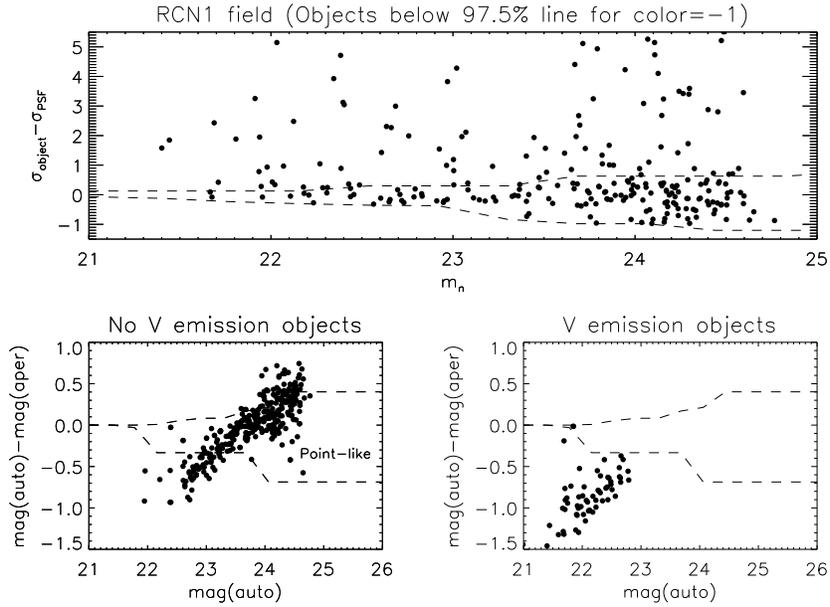}
\caption{Top panel: $\sigma_{object}-\sigma_{PSF}$ vs. $m_n$ magnitude
for catalogued emission objects. 
The dashed lines shows the $\sigma_{object}-\sigma_{PSF}$ boundaries 
for a simulated distribution of point-like objects. 
Bottom panel: mag(auto) - mag(aper) vs. mag(auto) of the objects selected in 
Figure~\ref{fig2}. The dashed lines show the boundaries within which the 
distribution of modelled point-like objects falls. 
In both panels point-like emission objects are those which lie between 
the dashed lines. Objects with no emission in V are both point-like and 
extended (lower left panel), whereas objects with luminosity in V are all 
extended (lower right panel). \label{fig3}}
\end{figure}

The stellar PSF for our images can be modelled with a 
two dimensional Gaussian. Twenty un-saturated stars were selected in 
the narrow band image and a 2D Gaussian fit was produced with the task 
FITPSF within IRAF.   
Our emission line candidates are faint, and their fluxes in the PSF wings 
looks as background noise; therefore we modeled the light distribution in our
selected emission line candidates using a 2D Gaussian. 
Those objects with $\sigma_{object}-\sigma_{PSF}>>0$ 
are most probably extended objects, while those with 
$\sigma_{object}-\sigma_{PSF}\sim 0$ are most likely
point like objects. Because our candidates are faint, 
we need to test whether noise and small changes of the 
PSF across the wide field can affect their light
distribution in a systematic way.
We simulated sources with $\sigma_{object}=\sigma_{PSF}$ and
[OIII] magnitude in the range from 20 to 25. 
These simulated objects were added onto the narrow band image
and a 2D Gaussian was fitted in the same way as for the real objects. 
The results are shown in Figure~\ref{fig3}; the dashed lines over plotted 
in the upper panel of Figure~\ref{fig3} show the maximum and minimum value of 
$\sigma_{object}-\sigma_{PSF}$ in each m$_n$ magnitude bin of
200 modelled objects. 
The range in $\sigma_{object}-\sigma_{PSF}$ is larger for fainter
objects, where the noise affects their 2D light distribution.  Based on
this simulation, point-like objects are those candidates whose
$\sigma_{object}-\sigma_{PSF}$ lies between the dashed lines in the
upper panel of Figure~\ref{fig3}. Those objects whose
$\sigma_{object}-\sigma_{PSF}$ happen to fall below the dashed-line
curve, turned out to be bad pixels or have cosmic-rays residual in the
area influenced by the CCD gaps, where fewer frames were averaged, and
were not plotted. They were rejected in the final catalogue.

\subsubsection{Point-like objects: magnitude selections}
An independent test on whether a source is extended or not
is based on the mag(auto)-mag(aper) vs. mag(auto) diagrams 
for the detected sources. In our extraction procedure
the aperture magnitude mag(aper) is computed in a fixed
aperture of diameter 1.5$''$, equivalent to 
$3\times \sigma$ of the mean seeing measured on the image. 
The so-called mag(auto) is the same to 0.1 mag as 
mag(aper) for point-like objects,
while the two magnitudes differ for extended objects. \\
For the point-like objects modelled in our Monte-Carlo simulations,
we studied the distribution of mag(auto)$-$mag(aper) 
as a function of the object magnitude mag(auto). 
This is shown in the lower panel of Figure~\ref{fig3}. 
As in the case of the 2D Gaussian fit, for bright point-like objects 
this difference is fixed and at faint magnitudes we have a larger spread
because of photometric errors. 
Point-like objects are those which lie between the dashed lines 
of Figure~\ref{fig3}.\\

The final set of point-like objects in our extracted catalogue 
is selected as those which satisfy both criteria: a 2D Gaussian fit
which is consistent with the PSF, and mag(auto)-mag(aper) 
inside the distribution for simulated point sources. 
This is a necessary condition for our selected objects to
be selected as point-like, but it is not sufficient, i.e.
high redshift starbursts with a strongly peaked central luminosity may
still get into our sample.

\subsection{Spectroscopic confirmation}
For spectroscopic validation, we use a sample of spectra obtained with
the 2dF fiber spectrometer at the 4 m Anglo Australian Telescope
by Freeman et al. (2000; 2001, in prep.).
Spectra were acquired with a spectral resolution R=2000 in the 
wavelength range 4550 \AA\ to 5650 \AA\ and a total integration time 
of 5 hrs on the 15th of March 1999.

\begin{figure}
\epsscale{0.7}
\plotone{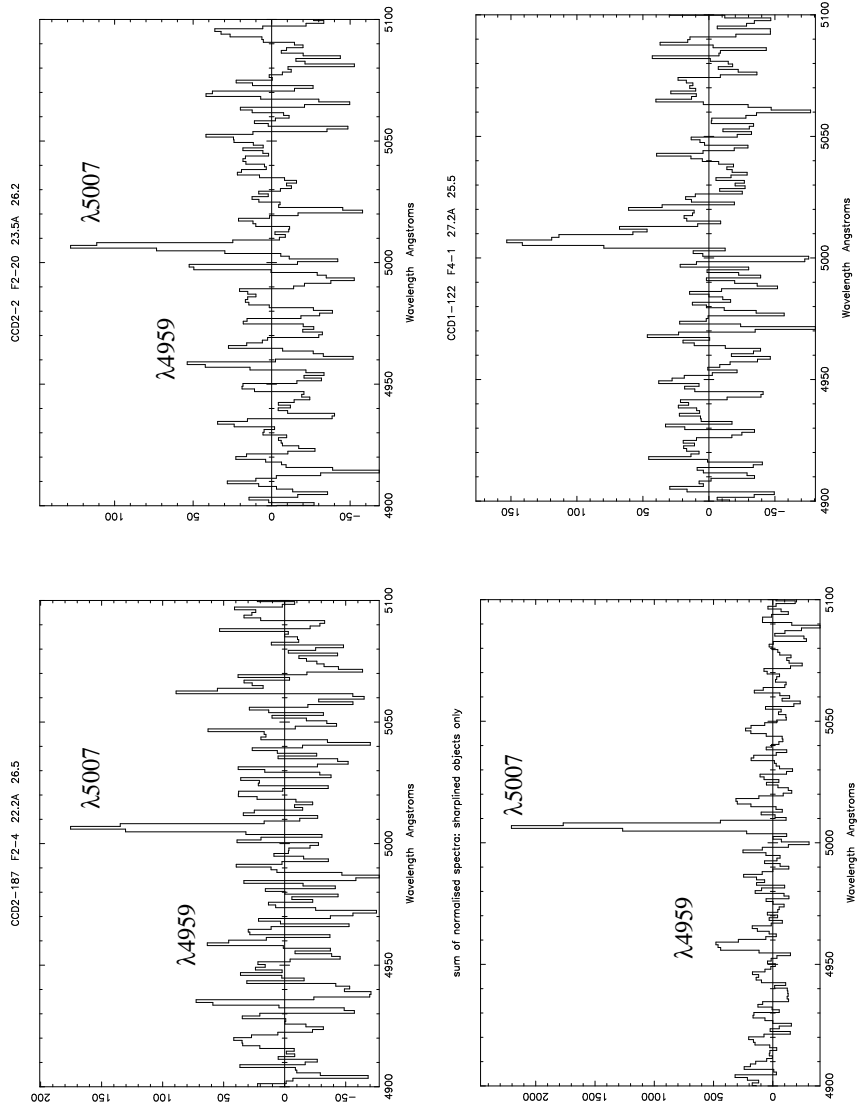}
\caption{Top row: spectra of two ICPNe candidate from the AAT 2dF run.
The 5007 \AA\ line is visible with S/N $> 3$ and a second line at
4959 \AA\, corresponds to the second emission line of the [OIII] doublet.
Lower left corner: cumulative spectrum of all 23 ICPNe from the 2dF run. 
Each spectrum was shifted to zero velocity and normalized by the 
flux in the 5007 \AA\ [OIII] line. The equivalenth width ratio of the [OIII]
5007\AA\ / 4959 \AA\ is $3.2 \pm 0.1$, so the real fraction of ICPNe in
this sample is $0.94 \pm 0.03$ (Freeman et al. 2000, 2001 in preparation). 
Lower right corner: the 2dF spectrum of a likely 
Ly$\alpha$ emitter. The asymmetric profile of the line is clearly visible.
\label{ICPNEspec}}
\end{figure}

A total of 43 fibers were allocated to the narrow-line emitter candidates 
selected by Feldmeier et al. 1998 at coordinate $\alpha (J2000) = 12:30:39\, 
\mbox{and}\, \delta (J2000) = 12:38:10$; we will refer to this field 
as the FCJ field. In this field, 15 objects were spectroscopically 
confirmed as emission line objects. 
The best S/N spectra of the ICPNe candidates in the FCJ field 
are shown in Figure~\ref{ICPNEspec}. As a further test on the intrinsic
nature of these objects, 23 spectra of ICPNe candidate from the whole
area in Virgo surveyed with the 2dF (see Freeman et al.
2000 for additional details), were summed: these 23
objects include the 15 from the FCJ field and another 8 from other  
Virgo intracluster empty fields. 
The cumulative spectrum shows the [OIII] doublet very clearly, 
with the expected equivalent width ratio $EW_{5007} = 3 \times EW_{4959}$
for the [OIII] doublet (see Figure~\ref{ICPNEspec}).
Of the 15  spectroscopically confirmed objects in the FCJ field, 
13 are ICPNe and 2 are broad-lined Ly$\alpha$ emitters.
The broad-lined sources were identified as Ly$\alpha$ by 
Freeman et al. (2000) and Kudritzki et al. (2000) because of their 
typical asymmetric line-profile; furthermore in Kudritzki et al. (2000) 
no H$\beta$ and [OIII] emissions were detected at the expected redshifted 
wavelengths, under the hyphothesis that the strongest emission was [OII] 
at redshift 0.35.

These observations show that a significant fraction of emission 
line candidates selected carefully with the ``on-off'' band
technique are indeed ICPNe. They were used to
make the first kinematical study of the ICPNe population in Virgo.
These issues will be discussed by Freeman et al. (2001, in preparation).
Here these objects are used for validation of the automatic 
selection procedure.

\paragraph{Validation of the selection procedure}
To compare our photometric selection of emission line objects with 
the spectroscopic confirmations, we have applied 
our selection procedure to the ``on-off'' images for the 
FCJ field.
The survey on the FCJ field was carried out with a 44 \AA\ narrow band 
[OIII] filter, with central wavelength at $\lambda_c = 5027$ \AA\footnote{The 
conversion from $m_{AB}$ to $m_{5007}$ for this filter is 
$m_{5007} = m_{AB} + 3.04$.}.  
Of the 43 emission line candidates in FCJ selected with the blinking
technique and allocated a fiber in the spectroscopic 
follow-up, we could detect and measure narrow band magnitudes in the 
images for 36 of them. 
The remaining 7 objects were very near to bright objects
and were therefore discarded because of likely errors in the SExtractor
photometry. Of these 7 objects, 4 were spectroscopically
confirmed (3 ICPNe + one Ly$\alpha$).
In the following, we discuss the percentage of spectroscopically confirmed
objects in the sample of 36 objects with allocated fiber
which are in common with our selected sample. 
In this sample, 11 objects (i.e. 30\%) are spectroscopically 
confirmed as emission line objects; of these objects 10 are confirmed ICPNe
and one is a Ly$\alpha$ candidate.  

\begin{figure}
\epsscale{0.7}
\plotone{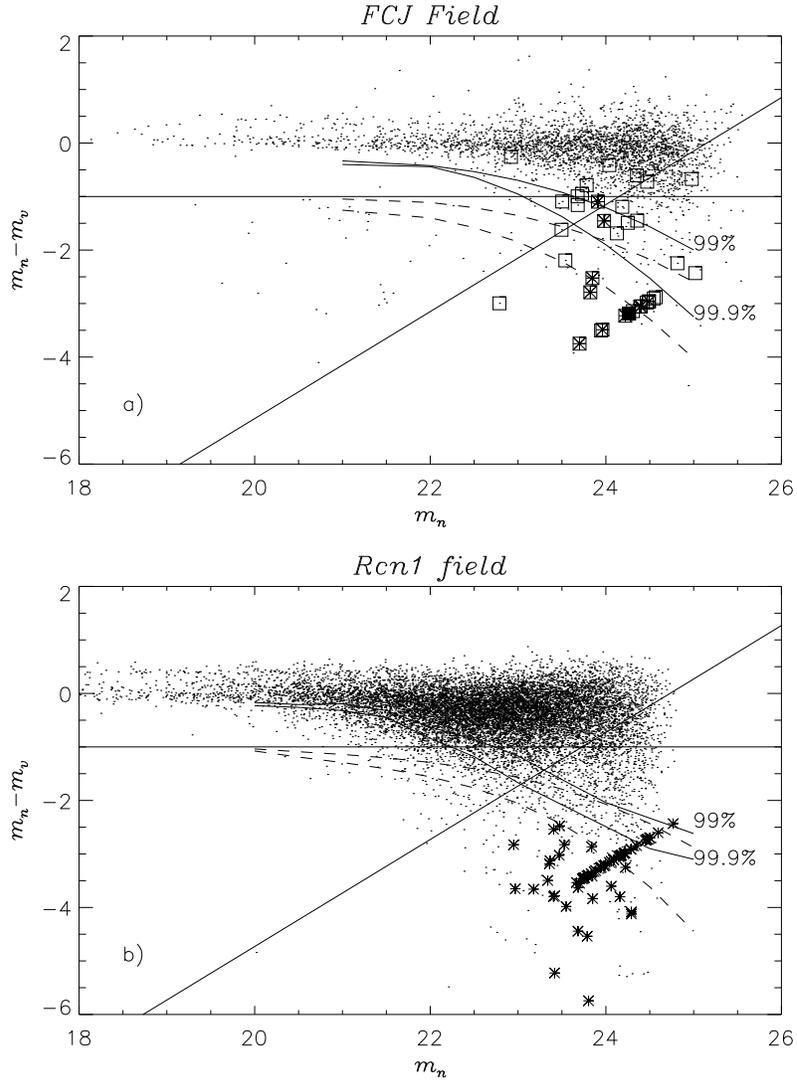}
\caption{Color magnitude diagrams for the FCJ  and the RCN1 fields.
Upper panel: CMD for the FCJ field. The lines have the same meaning as in 
Figure~\ref{fig2}.  
The diagonal line correspond to $m_v$ = 24.75. which is the 
$1.0 \times \sigma$ limiting magnitude in the V band.
The over-plotted squares represent 
the allocated fibers of Freeman et al. (2000, 2001, in prep.) 
and the asterisks are the spectroscopically confirmed ICPNe. 
The full square represents a 
spectroscopically confirmed Ly$\alpha$ emitter located in this field. 
See text for details \label{fig4}. Two spectroscopically confirmed 
PNe have off band fluxes near the limiting magnitude because of
scattered light from bright emitters in the field.
Fourteen objects in the Feldmeier et al. (1998) catalogue have no
computed mag(aper) in the off-band, assigned by SExtractor in dual image
mode: they have been assigned $m_v$ = 27.0 (see text).
Lower panel: CMD for the RCN1 field. The ICPNe candidates are indicated
with asterisks; those objects with no computed mag(aper) in the off-band, 
using SExtractor in dual image mode, have been assigned
$m_v$ = 27.0.}
\end{figure}

In Figure~\ref{fig4} we show the CMD for all sources 
in this field. Squares indicate those sources in common 
with Feldmeier's catalogue of emission line objects with an 
allocated fiber, while asterisks indicate those ICPNe candidates  
with confirmed spectra; the Ly$\alpha$ is indicated with a full square.
The lines plotted in Figure~\ref{fig4} have the same meaning as in 
Figure~\ref{fig2}. 
For 16 out of the 36 allocated fiber, SExtractor could
not detect a $m_v$ magnitude at the $x,y$ position of the [OIII]
detected emission. All these objects would be selected 
and are reported on this diagram assuming a $m_v$=27\footnote{
Here we adopted the V magnitude that a star would have
if the flux from a [OIII] emission of $m_{AB}=24.5$ would have been 
seen through a V band filter}.
Two out of the 10 confirmed ICPNe were in the region of
the frame with bright scattered light from nearby luminous objects.
So their V magnitude appear above the detection limit, because 
of scattered background light. 
The 8 out of the 10 confirmed ICPNe and the Ly$\alpha$ emitter are found
\begin{itemize}
\item below the line for detection in the off-filter,
\item below the $97.5\%$ line of objects with colour$= -1$\footnote{For the
FCJ filter, this corresponds to an EW$\sim 70 $\AA.}. 
\item or in the sample with no detected off-band mag(aper),
when SExtractor is used in dual image mode.
\end{itemize}
The sample selected according to these criteria has a higher spectroscopic 
confirmation rate, 50\%, compared to the Feldmeier's sample for the 
FCJ field, i.e. 30\%. 
The part of the 36 objects which do not satisfy the above criteria
have a much lower spectroscopic confirmation rate: 2/18 = 10\%.
These percentages are lower limits because,
as Fig~\ref{ICPNEspec} shows, they are based on low signal-to-noise spectra.

We also ran the morphological selection procedure on the spectroscopically 
confirmed emission line objects, which are in common with our
photometrically selected sample. 
Eight objects which showed an unresolved [OIII] $\lambda 5007\AA$ emission 
and the additional $\lambda 4959\AA$ line are point-like sources. 
These are the confirmed ICPNe.  
One object which is spectroscopically confirmed
as an emission line object with a broad asymmetric emission line (see
Figure~\ref{ICPNEspec}) is classified as extended, and it is probably 
a Ly$\alpha$ emitter.

Based on the spectroscopically confirmed sample, our selection criteria 
for Bona Fide ICPNe candidates are the following: these must be
point-like objects lying below the 97.5\% line for intrinsic color 
$m_n - m_v <-1$, and with no detection in the off-band (V) image.
This is the population of objects that we discuss hereafter in the RCN1
field. From Figure~\ref{fig3}, we found 75 point-like objects 
with those characteristics. These are the ICPNe candidate in our RCN1 field.

\section{The Virgo cluster ICPNe}
The observations presented in this work allow us to obtain 
a large statistical sample of IPNe candidates in the RCN1 field.
Using also the data from the FCJ field, we can make a preliminary 
study of their distribution on the sky. 

\subsection{Luminosity function and cluster depth}
The PNLF has been used extensively as a distance indicator. 
Observations in ellipticals, spirals and irregular galaxies have shown 
a PNLF truncated at the bright end. 
The shape of this LF is given by the 
semi-empirical fit:
\begin{equation}
N(M) = c_{1} e^{c_{2}M} [1-e^{3(M^{*}-M)}]\label{PNLF}
\end{equation}
where $c_{1}$ is a positive constant, $c_{2}= 0.307$ and 
the cut-off 
$M^{*}(5007)=-4.5$ \cite{cia89}. The evidence for 
a bright cut-off comes from the observations of nearby
galaxies, and its dependence on the age and metallicity  
of the stellar population was analysed in some detail by \citep{dopi92,me93}. 
Its overall metallicity dependence seems to be modest provided that the 
metallicity 
differences are no more than about 30\%, but see also M\'endez et al. (1993) 
for an elaborate discussion of the effects of sample size and population age.

\begin{figure}
\epsscale{0.7}
\plotone{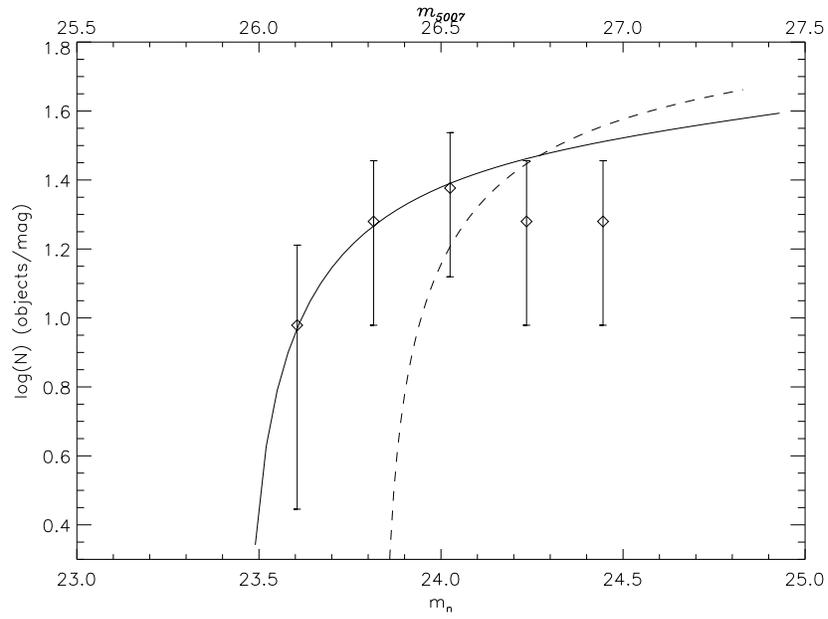}
\caption{Luminosity function of the spectroscopically confirmed ICPNe 
from FCJ field. The continuous line is the best fit for a 
distance modulus equal to 30.53. The dashed lines indicates the same
PNLF for the distance modulus 30.86 assigned to M87 using the 
PNLF within 20 arcmin of M87.\label{fig5}}
\end{figure}

Figure~\ref{fig5} shows the luminosity function (LF) of the
spectroscopically confirmed ICPNe from the FCJ field, with the best fit c1 and 
distance modulus values computed for a single population of PNe at a 
fixed distance\footnote{The $m_{5007}$ magnitudes are from 
Feldmeier's 1998 data.}, using a $\chi^2$ minimization test. 
The fit was done in the first magnitude interval (first three bins) 
where the sample is complete.  The best-fit distance modulus is 30.53, 
once the empirical PNLF is convolved with the photometric errors.  
On the same plot, we show the PNLF which is expected for a distance 
modulus of 30.86, as determined from PNe in the inner part of M87 
\cite{cia98}.  Because metallicity and age effects do not strongly 
affect the bright cut-off \cite{dopi92}, the most probable interpretation 
of Figure~\ref{fig5} is that the spectroscopically confirmed ICPNe 
in the FCJ field are closer 
to us than the M87 PNe.  When a large area is surveyed in Virgo empty fields,
we preferentially pick up those ICPNe which are nearer to us,
i.e. brighter.  Thus the inferred front edge of the ICPN population in
the FCJ field is at 14\% shorter distance than M87, which 
corresponds to 2.1 Mpc, when the distance to M87 is
at 14.9 Mpc.  Ciardullo et al. (1998) and Feldmeier et al. (1998) 
already suggested that the LF of ICPNe may be distorted because of the 
Virgo cluster depth.  However, we have now spectra which confirm 
this hypothesis.

\begin{figure}
\epsscale{0.7}
\plotone{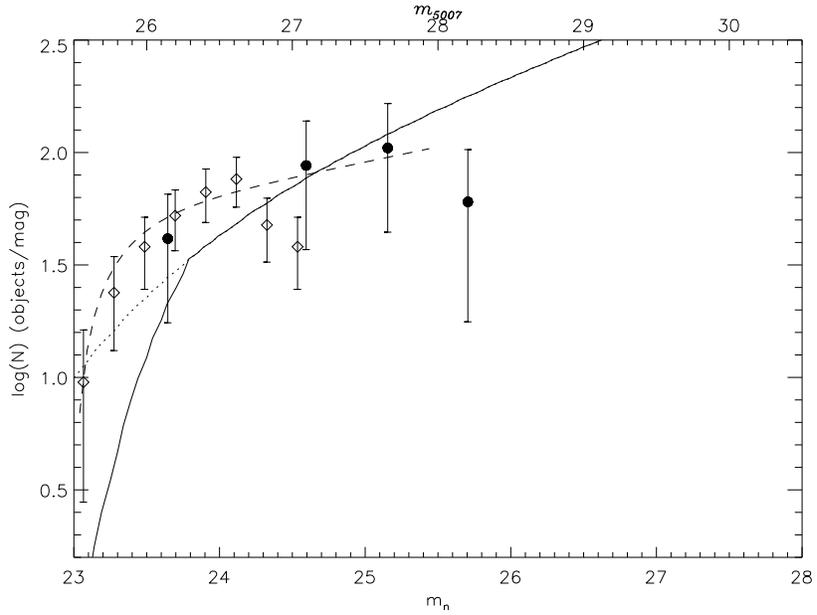}
\caption{Luminosity function of the ICPNe sample selected in the RCN1 field. 
The dashed line shows the PNLF for a distance modulus of 30.29, convolved
with the photometric errors. On the same
plot the continuos line shows the expected LF in our narrow filter for the
Ly$\alpha$ population in the field at redshift z= 3.13 from Steidel et al.\
(2000), see discussion in the text. The faint dotted line shows the expected 
contribution from the Ly$\alpha$ emitters with V magnitudes brigther than 
24.75. Full dots indicate the LF in our narrow band filter from Ly$\alpha$
emitters in the blank-field survey by Cowie \& Hu (1998).
\label{fig6}}\end{figure}

Figure~\ref{fig6} shows the PNLF for the ICPNe candidates from the
RCN1 field. Here again the cut-off is at brighter magnitudes than for
the PNe in M87.  Could this be due to contamination by Ly$\alpha$
emitting objects at $z \sim 3.13$, which are also picked up by our
technique?  
There is no well-sampled Ly$\alpha$ LF of such objects available in
the literature. We have therefore estimated the LF of Ly$\alpha$
emitters in our sample from the work of Steidel et al.\ (2000). These
authors showed that the continuum LFs of Ly$\alpha$ emission selected
galaxies and of field Lyman break galaxies (LBGs) are consistent with
being the same, and that there is no evidence for enhanced Ly$\alpha$
emission from fainter galaxies. The equivalent width (EW)
distribution of their spectroscopic sample is consistent with being a
Gaussian centered at zero, such that the fraction of objects with
EW$>80$\AA\ (restframe EW$>20$\AA) is $\simeq 25\%$. We can thus estimate
the narrow band magnitude LF of Ly$\alpha$ emitters in our on-band
filter as follows. We take the LF of field LBGs from Steidel et al.\ 
(1999) with the parameters given in their Fig.~8 except m$_\ast^{\rm V} 
= {\rm m}_\ast^{\rm R} + 0.24$ (as LBGs have about $20\%$
fainter flux density in V than in R; M. Giavalisco, private
communication). We then apply the same selection criteria
to those sources as to our ICPNe candidates, that m$_{\rm V}=24.75$, 
and finally convolve with the EW distribution for the Ly$\alpha$ 
emission, accepting only those galaxies
with an observed EW$> 110$\AA\ which correspond to our sample selection 
criteria. The resulting LF, scaled to our larger field and effective
volume, is shown as the full line in Fig.~\ref{fig6}.

We have also added the observed data points constructed from the
Ly$\alpha$ blank field search done by Cowie \& Hu (1998), as this
survey is carried out in a very similar way to our survey for ICPNe in
the Virgo cluster fields. We used their V magnitudes and 
EWs listed in their Table~1 for those objects whose V
magnitudes are fainter then 24.75, and then converted to our filter AB
magnitudes. Once the resulting LF is scaled to our larger field, the
Cowie \& Hu (1998) narrow band LF for the Ly$\alpha$ emitters agrees
very well with that computed from Steidel et al. (2000).  The
comparison of the candidate ICPNe LF in the RCN1 field with the
computed narrow band LF of the field Ly$\alpha$ emitters from Steidel
et al. (2000) and Cowie \& Hu (1998) show that the Ly$\alpha$
contribution does not dominate in the narrow band AB magnitude range
23-24: the fraction of expected contamination by Ly$\alpha$ emitters
with V magnitude fainter of 24.75 in the first magnitude bin is 15\%.
This is in agreement (within the errors) with the fraction of
Ly$\alpha$ contaminants of 25\%, as obtained from the spectroscopic
sample

Therefore most over-luminous bright objects in the RCN1 are likely 
to be ICPNe, and they are brighter than those at the cut-off in the 
M87 PNLF because they too are closer to us than M87. 
Once we have convolved the empirical PNLF given by Equation~\ref{PNLF} with
the photometric error of our sample, we fit the data point  
of RCN1 LF and obtain the best fit distance modulus of $30.29$
with the following errors $^{+0.2}_{-0.2}$, which refers to the 90\% c.l..
Our estimate of the distance modulus is slightly biased from 
the contribution of the [OIII] emission at 4958.9 \AA\, which enters
in our filter band pass for radial velocities $v_{rad} > 1580$ km s$^{-1}$.
We have estimated this systematic contribution assuming a Gaussian
velocity distribution for the ICPNe in Virgo (with $v_{mean} = 1100$
km s$^{-1}$ and $\sigma = 800$ km s$^{-1}$). This gives a maximal
brightening of the PNLF of 0.13 mag, assuming that most ICPNe
at the front edge of the cluster have the same velocity dispersion
as galaxies in Virgo.
If this systematic brightening is taken into account, the distance modulus to
the front edge of the Virgo cluster is 30.4$\pm$0.2 at RCN1, in agreement with the 
distance modulus derived for the spectroscopic sample, within the
photometric errors.
This implies that the front edge population seen in the RCN1 Virgo field is
at 19\% shorter distance than M87, i.e. 2.8 Mpc in front of the cluster
center. The Virgo cluster appears to have a significant depth.

Previously, Yasuda et al. (1997) have argued that the distribution of
spiral galaxies in Virgo is best described as an elongated, cigar-like
structure with a depth $\pm 4$ Mpc from M87. Neilsen \& Tsvetanov
(2000) and West \& Blakeslee (2000) have used surface brightness
fluctuations to determine the distances to the Virgo ellipticals, and
found that most of these are arranged in a nearly collinear structure
extending about $\pm (2-3)$ Mpc from M87. Therefore our result from the
ICPNe population is not entirely surprising.

\subsection{Luminosity surface density}
From our results we can determine a surface brightness associated with
the diffuse population of evolved stars in the FCJ and the RCN1
fields. Given our tight selection criteria in the photometric analysis
and the likely incompleteness of the 2dF detections, this will in both fields
be a lower limit.   
From stellar evolution theory, it is expected that the 
luminosity-specific stellar death rate of a galaxy should be independent of 
the precise state of the underlying stellar population. 
\cite{rebu86} have shown that this quantity is remarkably independent 
of both age and initial mass function: if this is the case, the 
number density of planetaries per unit bolometric luminosity should be 
approximately the same for every galaxy.
Here we adopt the luminosity-specific planetary nebulae
density from M31, $\alpha_{1,B} = 9.4 \times 10^{-9}$ PNe $~
L_{B\odot}^{-1}$ \cite{cia89} to infer the amount of light
associated with the ICPNe.\footnote{The most appropriate value 
of $\alpha_{1,B}$ for the intracluster light is unknown at this stage,
therefore we adopted the $\alpha_{1,B}$ for the bulge of M31, which is an evolved
population similar to those in early-type galaxies. Independent measurements
of $\alpha_{1,B}$ in ellipticals shows a variation up to a factor 5.} 
For the spectroscopically confirmed sample in the FCJ field, 
the inferred luminosity surface density is
$5.2\times 10^6~L_{B\odot}{\rm arcmin}^{-2}$, corresponding to
$\simeq 0.28~L_{B\odot}$ pc$^{-2}$ at a distance of $\simeq 15$ Mpc.
For the RCN1 field, we have a sample of 75 candidates.  From the
spectroscopic follow-up by Freeman et al. (2000) the contamination by high
redshift objects can still be up to 25\%, and the possibility of large-scale
structures in the Ly$\alpha$ emitters implies some uncertainties in a 
simple statistical subtraction for ICPN measurements. Removing $25\%$ of our sources 
gives a likely lower limit to the surface luminosity density of the
ICPNe population. Thus we take the number of ICPNe in this field to be
56, which on an area of $31.8 \times 30.1 $ arcmin$^2$ amounts to a
luminosity surface density of $6.2\times 10^6 L_{B\odot}{\rm
arcmin}^{-2}$, which corresponds to $\simeq 0.33 L_{B\odot}$pc$^{-2}$, 
similar to that for the spectroscopically selected sample. The
inferred blue surface brightness is $\mu_B = 28$ m$_B$ arcsec$^{-2}$.
The inferred luminosity derived assuming all the sample at a single distance
gives the smallest amount of intracluster light: from Feldmeier 
et al. (1998), if the ICPN were spherically distributed with a 
constant density, the amount of starlight could be up to a factor of 
three larger.

The result that the inferred intracluster surface brightnesses for both
fields are similar is robust and very interesting, because the FCJ field 
is $45'$
from the center of the cluster core, while the RCN1 field is 1.5 deg
from the center.  The azimuthally averaged total luminosity density
contributed by galaxies is given by Binggeli et al. (1987) as a function of
distance to the core center.  This function has a large scatter (see
Figure~16 \cite{bin87}), and implies $\sim 10^{11}$ L$_{B\odot}$
deg$^{-2}$ at the FCJ distance ($45'$) and $\sim 3\cdot10^{10}$
L$_{B\odot}$ deg$^{-2}$ at the RCN1 distance (1.5 deg).

Over the range of radii probed by these two fields, the luminosity
surface density of galaxies in Virgo decreases by a factor of $\sim 3$
\cite{bin87} while that for the ICPNe population is nearly
constant. The diffuse light contributes from 17\% (FCJ) to 43\%
(RCN1) of the total (i.e. galaxies + diffuse light) luminosity density 
in these two fields (recall that these numbers are lower limits from our data).
Because we have no evidence for a radial gradient in the ICPNe
luminosity profile, it is very difficult to estimate the total
luminosity for this population at this stage.  

\section{Conclusions}
We have imaged an empty area of $34\times 34$ arcmin$^2$ in the Virgo
cluster in a narrow band filter centred at the redshifted Virgo [OIII]
emission and in the V band.  The area surveyed in this work is the
largest probed for ICPNe in one single pointing so far, and it allows us to 
get a better statistical sample of ICPNe.  We have developed an automatic
extraction procedure for ICPNe candidates using publicly available
photometry software (SExtractor).  This selection procedure was
validated using a spectroscopic confirmed sample of ICPNe. In our
field 75 objects were found which satisfy the selection criteria for
ICPNe, i.e., point-like, no measurable continuum, and equivalent
width EW $> 110$\AA\ after error convolution.

The luminosity functions for both the spectroscopically selected
sample and the ICPNe sample in the new field show a brighter cut-off
compared to the M87 PNe. We have evaluated the possible contribution
of Ly$\alpha$ emitting objects at redshift $z=3.13$ from
Steidel et al. (2000) and found that these objects contribute mostly at
1-2 magnitudes fainter than the bright cutoff found here, and in any
case only up to 25\% of the total sample. The bright cutoff is
therefore most probably due to the three-dimensional depth of the
Virgo cluster: our results place it at 14\% - 19\% shorter distance
than M87. Our selected sample traces preferentially the edge of
the cluster near us.

The luminosity surface density in these two fields is about
$6 \times 10^6 L_{B\odot} $ arcmin$^{-2}$ with both fields
having similar densities. The lack of radial gradient between
both fields implies that the total amount of luminosity in the
diffuse component is very sensitive to the outer cut-off, and
could be comparable to the luminosity in galaxies.

\acknowledgments The authors wish to thank the referee, Dr. R. Ciardullo,
for his insightful comments. Moreover we would like to thank the ESO 2.2m
telescope team for their help and support during the observations
carried out for this project, in particular E. Pompei and H. Jones.
We thank F. Valdes for his help and support in using MSCRED in
reducing WFI mosaic data, the OACDF team for support during astrometry
and co-addition, and E. Bertin for his help and suggestions in using
SExtractor.  The authors acknowledge the use of NED and the CDS public
catalogues.  JALA and OG were supported by the grant 20-56888.99 from
the Schweizerischer Nationalfonds. NRN was supported by the European 
Social Funds and acknowledges the support from the University Federico II 
for travel grants within the program for International Exchange.

\appendix

\section{m(5007) and the AB magnitude system}
Absolute flux calibrations for our narrow band imaging (with an
interference filter centred on the [OIII] redshifted wavelength at the 
Virgo systemic velocity) were performed
according to Jacoby et al. (1987). These calibrations 
give us the total flux in the [OIII] line in ergs sec$^{-1}$ cm$^{-2}$,
which we indicate as $F_{5007}$. 
Following Jacoby (1989), the m(5007) magnitudes are defined as:
$$
m(5007) = - 2.5\log F_{5007} - 13.74
$$
The magnitude in the AB system are given by:
$$
m_{AB} = -2.5(\log F_\nu + 19.436) 
$$
and we can rewrite m$_{AB}$ as function of m(5007) as follows:
$$
m_{AB} = -2.5\log F_{5007} -2.5\log
 \left({\lambda_c^2\over\Delta\lambda_{eff}c}\right) - 48.59
= m(5007) - 2.51  
$$
where
$$
\Delta \lambda_{eff} = \int d\lambda T(\lambda),
$$
$\lambda_c$ is the filter central wavelength, and $T(\lambda)$
is the filter bandpass.


\clearpage



\clearpage

\begin{deluxetable}{ccc}
\tablecolumns{3} 
\tablewidth{0pc} 
\tablecaption{Number of modelled point-like objects and spurious detections
recovered from simulations.}
\tablehead{
\colhead{LOWTHRESHOLD}    &  \colhead{Point-like objects}  & \colhead{Spurious detections}}
\startdata
0.7 $\sigma$ &  680  & 1001  \\
0.8 $\sigma$ &  567  &237   \\
0.9 $\sigma$ &  478  & 66    \\
1.0 $\sigma$ &  410  & 30    \\
1.1 $\sigma$ &  362  & 16    \\
1.2 $\sigma$ &  310  & 11    \\
\enddata 
\end{deluxetable}


\begin{thebibliography}{}

\bibitem[Arnaboldi et al. 1996]{aa96} Arnaboldi, M., et al. 1996, \apj, 472, 145

\bibitem[Bernstein et al. 1995]{ber95} Bernstein, G.M. et al. 1995, AJ, 110, 
1507

\bibitem[Bertin \& Arnouts 1996]{ber96} Bertin, E. \& Arnouts, S. 1996, A\&AS
 117, 3993 

\bibitem[Binggeli et al. 1987]{bin87} Binggeli, B., Tammann, G.A., 
Sandage, A. 1987, AJ, 94, 251

\bibitem[Calc\'aneo-Rold\'an et al. 2000]{calc00} Calc\'aneo-Rold\'an, C., 
Moore, B., Bland-Hawthorn, J., Malin, D., Sadler, E.M. 2000, MNRAS, 314, 324

\bibitem[Capaccioli et al. 2001]{cap2001} Capaccioli, M., et al. 2001, A\&A,
submitted

\bibitem[Ciardullo et al. 1989]{cia89} Ciardullo, R., Jacoby, G.H., Ford, 
H.C., Neill, J.D. 1989, \apj, 339, 53

\bibitem[Ciardullo et al. 1998]{cia98} Ciardullo, R., Jacoby, G.H., 
Feldmeier, J.J., Barlett, R. 1998, \apj, 492, 62

\bibitem[Colless et al. 1990]{col90} Colless, M., Ellis, R.S., 
Taylor, K., Hook, R.N. 1990, MNRAS 244, 408 

\bibitem[Cowie \& Hu 1998]{cohu98} Cowie, L.L., \& Hu, E.M. 1998, AJ, 115, 1319

\bibitem[Crane et al. 1977]{crtaw77} Crane, P., Tammann, G.A., Woltjer, L.
1977, Nature, 265, 124

\bibitem[Dopita et al. 1992]{dopi92} Dopita, M.A., Jacoby, G.H., Vassiliadis, 
E. 1992, \apj, 389, 27

\bibitem[Durrell et al. 2001]{durre01} Durrell, P., Ciardullo, R., Feldmeier, 
J., Jacoby, G.H. 2001, \apj, submitted 

\bibitem[Feldmeier et al. 1998]{fel98} Feldmeier, J.J., Ciardullo, R., 
Jacoby, G.H. 1998, \apj, 503, 109

\bibitem[Feldmeier 2000]{fel00} Feldmeier, J.J. 2000, Ph D. Thesis, 
Penn State University

\bibitem[Ferguson et al. 1998]{fer98} Ferguson, H., Tanvir, N.R., von Hippel, 
T. 1998, Nature, 391, 461

\bibitem[Freeman et al. 2000]{kcf2000} Freeman, K.C., et al. 2000, ASP Conf. 
Series 197, 389

\bibitem[Hammer et al. 1997]{} Hammer, F., et al. 1997, \apj, 481, 49 

\bibitem[Hogg et al. 1998]{} Hogg, D.W., Cohen, J. G., Blandford, R.,
Pahre, M. A. 1998, \apj, 504, 622 

\bibitem[Hui et al. 1993]{hui93} Hui, X., Ford, H.C., Ciardullo, R., Jacoby, 
G.H. 1993, \apj 414, 463

\bibitem[Guldheus 1989]{guld89} Guldheus, D.H. 1989, \apj, 340, 661

\bibitem[Kron 1980]{kron80} Kron, R.G. 1980, \apj 343, 305

\bibitem[Kudritzki et al. 2000]{kud2000} Kudritzki, R.P., et al. 2000, \apj 536,
19 

\bibitem[Jacoby et al. 1987]{jac87} Jacoby, G.H., Quigley, R.J., \& Africano, 
J.L. 1987, PASP, 99, 672

\bibitem[Jacoby 1989]{jac89} Jacoby, G.H. 1989, \apj, 339, 39

\bibitem[Jacoby et al. 1990]{jac90} Jacoby, G., Ciardullo, R, Ford, H. 1990,
\apj, 356, 332

\bibitem[McMillan et al. 1993]{mc93} McMillan, R., Ciardullo, R., Jacoby, 
G.H. 1993, \apj, 416, 62

\bibitem[Melnick et al. 1977]{mel77} Melnick, J., White, S.D.M., Hoessel, J. 
1977, MNRAS, 180, 207 

\bibitem[M\'endez et al. 1993]{me93} M\'{e}ndez, R.H., Kudritzki, R.P., 
Ciardullo, R., Jacoby, G.H. 1993, A\&A , 275, 534

\bibitem[M\'endez et al. 1997]{me97} M\'{e}ndez, R.H. et al. 1997, ApJ, 491, L23

\bibitem[M\'endez et al. 2001]{me2001} M\'{e}ndez, R.H., Riffeser, A., 
Kudritzki, R.P.  et al. 2001, \apj, in press


\bibitem[Neilsen \& Tsvetanov 2000]{neil2000} Neilsen, E.H., Tsvetanov, Z.I.
2000, \apj, 536, 255


\bibitem[Oke 1990]{vega}Oke, J.B. 1990, AJ, 99, 1621

\bibitem[Pierce \& Tully 1988]{pietu88} Pierce, M., Tully, R.B. 1988, \apj,
330, 579

\bibitem[Renzini \& Buzzoni 1986]{rebu86} Renzini, A., Buzzoni, A. 1986, 
in Spectral Evolution of Galaxies, ed. C. Chiosi and A. Renzini 
(Dordrecht:Reidel), 195

\bibitem[Smith 1981]{sm81} Smith, H.A. 1981, AJ 86, 998 

\bibitem[Steidel et al. 1999]{lbg1999} Steidel, C.C., Adelberger, K.L., 
Giavalisco, M., Dickinson, M., Pettini, M. 1999, \apj, 519, 1

\bibitem[Steidel et al. 2000]{ly2000} Steidel, C.C., Adelberger, K.L., 
Shapley, A.E., Pettini, M., Dickinson, M., Giavalisco, M. 2000, \apj, 532, 170

\bibitem[Teplitz et al. 2000]{tep2000} Teplitz, H.I. et al. 2000, \apj, 542,
18

\bibitem[Theuns \& Warren 1997]{theu97} Theuns, T., Warren, S. 1997, MNRAS, 
284, L11

\bibitem[Thuan \& Kormendy 1977]{thuan77} Thuan, T.X., Kormendy, J. 1977, PASP,
 89, 466

\bibitem[Tonry et al. 1990]{tal1990} Tonry, J.L., Ajhar, E.A., Luppino, G.A. 1990,
AJ, 100, 1416

\bibitem[Uson et al. 1991]{uson91} Uson, J.M., Boughn, S.P., Kuhn, J.R. 1991, 
\apj, 369, 46

\bibitem[Vilchez-G\'omez et al. 1994]{vic94} Vilchez-G\'omez, R. Pell\'o, R., 
Sanahujia, B. 1994, A\&A, 283, 37

\bibitem[Welch \& Sastry 1971]{welch71} Welch, G.A., Sastry, G.N. 1971, \apj, 
169, 13 


\bibitem[West \& Blakeslee 2000]{west2000} West, M.J., Blakeslee, J.P. 2000,
ApJ, 543, L27


\bibitem[Yasuda et al. 1997]{yasu97} Yasuda, N., Fukugita, M., Okamura, S. 
1997, ApJSS, 108, 417

\bibitem[Zwicky 1951]{JL1.2} Zwicky, F. 1951, PASP, 63, 61

\end{thebibliography}
\end{document}